\documentclass[a4paper]{ESLAB2005}
\usepackage{epsfig}
\def\sax{{\em BeppoSAX\/}}

\begin{document}

\title{Exploring the Hard X-/soft gamma-ray Continuum Spectra with Laue Lenses}

\author{F. Frontera\inst{1,2} 
	\and A. Pisa\inst{1}
	\and P. De Chiara\inst{1}
	\and G. Loffredo\inst{1}  
	\and D. Pellicciotta\inst{1} 
	\and V. Carassiti\inst{3}
	\and F. Evangelisti\inst{3}
	\and K. Andersen\inst{4} 
	\and P. Courtois\inst{4} 
	\and B. Hamelin\inst{4} 
	\and L. Amati\inst{2}
	\and N. Auricchio\inst{2} 
	\and L. Bassani\inst{2}         
	\and E. Caroli\inst{2}
	\and G. Landini\inst{2} 
	\and M. Orlandini\inst{2} 
	\and J.B. Stephen\inst{2} 
	\and A. Comastri\inst{5} 
	\and J. Kn\"{o}dlseder\inst{6}
	\and P. von Ballmoos\inst{6}
} 

\institute{University of Ferrara, Dept of Physics, Ferrara, Italy
	\and IASF-INAF, Bologna, Italy
	\and INFN, Section of Ferrara, Italy
	\and Institute Laue--Langevin, Grenoble, France
	\and OAB, INAF, Bologna, Italy	
	\and CESR, Saclay, France
}

\maketitle 

\begin{abstract}
The history of X--ray astronomy has shown that any advancement in our 
knowledge of the X--ray sky  is strictly related to an increase 
in instrument sensitivity. At energies above 60 keV, there are interesting 
prospects for greatly improving the limiting sensitivity of the current 
generation of direct viewing telescopes (with or without coded masks), 
offered by the use of Laue lenses. We will discuss below the development 
status of a Hard X-Ray focusing Telescope (HAXTEL) based  on Laue 
lenses with a broad bandpass (from 60 to 600 keV) for the study of 
the X-ray continuum of celestial sources.  We show two examples
of  multi-lens configurations with expected sensitivity   
orders of magnitude better ($\sim 1 \times 10^{-8}$
photons~cm$^{-2}$~s$^{-1}$~keV$^{-1}$  at 200 keV) than that achieved 
so far. With this unprecedented 
sensitivity, very exciting astrophysical prospects  are opened.

\end{abstract}

\section{Introduction}
\label{intro}
\begin{figure}[ht]
 \begin{center}
    \epsfig{file=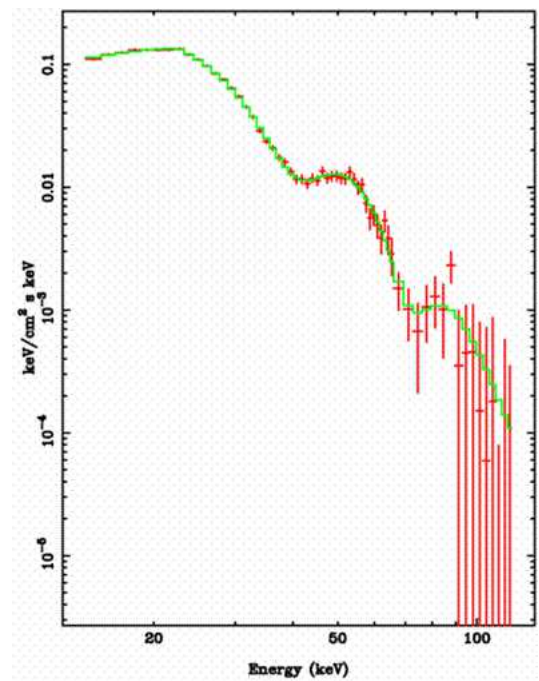, width=7cm}
  \end{center}
\caption[]{An example of the pulse-phase resolved spectrum of Her X--1 (corresponding to the fall of
the 1.24 s pulse profile), obtained with \sax\,
 soon after an anomalous low--ON cycle which occurred in October 2000.  The
second harmonics of the cyclotron line is apparent, as is  the low statistical quality of
the spectrum above 70 keV. Reprinted from \cite*{Orlandini05}. \label{f:herx1}}
\end{figure}

\begin{figure}[ht]
  \begin{center}
    \epsfig{file=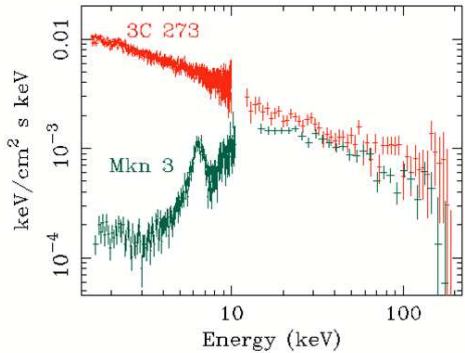, width=7cm}
  \end{center}
\caption[]{The energy spectrum, obtained with \sax\, of the Seyfert 2 galaxy MKN~3 
compared with that of the well known quasar 3C273. The peculiarity of
MKN~3 (Compton thick AGN) with respect to 3C273 (unabsorbed AGN) is apparent, as is the low
statistical quality of the two spectra above 100 keV. Figure kindly provided by M. Cappi.\label{f:mkn3}}
\end{figure}

Energy spectra beyond 60 keV are now well determined only for the hardest and 
strongest sources in the sky. For most objects, even for the most powerful, 
above 70-80 keV the spectra are scarcely known (see examples in Fig.~\ref{f:herx1},
Fig.~\ref{f:mkn3}). 

%
%

%

\begin{figure}[ht]
  \begin{center}
    \epsfig{file=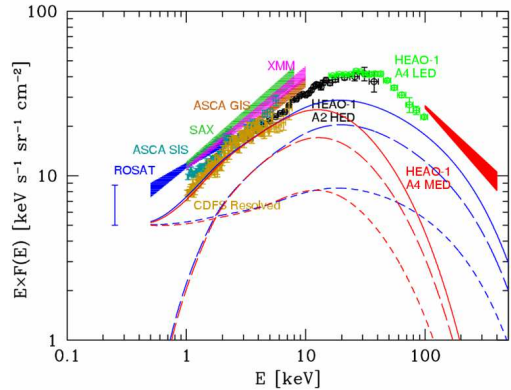, width=7cm}
  \end{center}
\caption[]{Measurements of the CXB $EF(E)$ spectrum
compared with a syntesis model (\cite{Comastri04}). In this model
it is assumed that the CXB is due to the superposition of Compton thin
($22<\log N_{\rm H} < 24$, long-dashed curves) and unabsorbed ($\log N_{\rm H}<22$,
short--dashed) AGNs, with two possible  values of the 
high energy cut-off: 400 keV (blue lines), 100 keV (red lines). The model parameters 
are tuned in order that their superpositions (continuous lines) can account 
for 80\% of the XMM (2-10 keV) measured CXB.\label{f:cxb}}
\end{figure}

However the  $>$60 keV energy channel  is crucial to study many open issues of 
high energy astrophysics. Among them, we wish to mention the following:
\begin{itemize}
\item
 The physics in  presence of super-strong magnetic fields 
(mass accreting X-ray pulsars, anomalous X-ray pulsars, Soft Gamma--Ray Repeaters);
\item
High energy cut-offs of BH binaries and, more generally, high energy tails of compact
X--ray binaries (low and high mass);
\item
Study of the hard X--/soft gamma-ray emission level from low and high mass X--ray binaries
in quiescence; 
\item
The role of Inverse Compton  with respect to synchrotron or thermal processes in GRBs.
\item
 The role of non thermal mechanisms in extended objects (supernova remnants, galaxy clusters).
\item
 High energy cut-offs in the spectra of Active Galactic Nuclei (AGN).
\item
 The origin of the high energy Cosmic X-/gamma-ray Background (CXB). 
Synthesis models (see, e.g., Fig.~\ref{f:cxb}) require a spectral roll-over 
with an e-folding energy of 
100-400 keV in AGN. So far only a few sparse measurements of these sources, 
mainly with \sax\, are available. It cannot be excluded that a new source population, with
a cut-off clustered around 100-400 keV, is responsible for the observed CXB
spectrum. Much more sensitive observations than those
performed with \sax\  at hard X--ray energies are urgent.
 \item 
Nuclear  (e.g., $^{44}$Ti) and annihilation (511 keV) lines.
\end{itemize}

A more extended discussion of these issues was given in the proposal submitted
in response to the ESA call for ideas for Cosmic Vision 2015--2025 "Exploring
the hard X--/gamma--ray continuum sky at unprecedented sensitivity" by Frontera
et al.).

%
%

\section{Development of high energy ($>$60 keV) lenses}
\label{s:lens}

\begin{figure}[ht]
  \begin{center}
    \epsfig{file=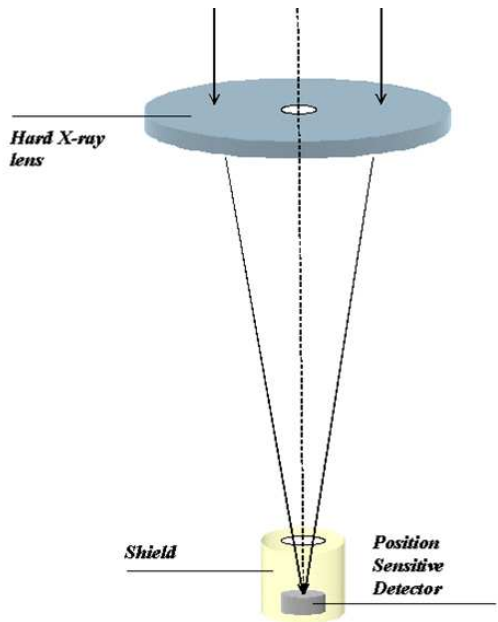, width=7cm}
  \end{center}
\caption[]{Pictorial sketch of a Laue lens. Reprinted from \cite*{Pisa04}.\label{f:lens}}
\end{figure}

\begin{figure}[ht]
  \begin{center}
    \epsfig{file=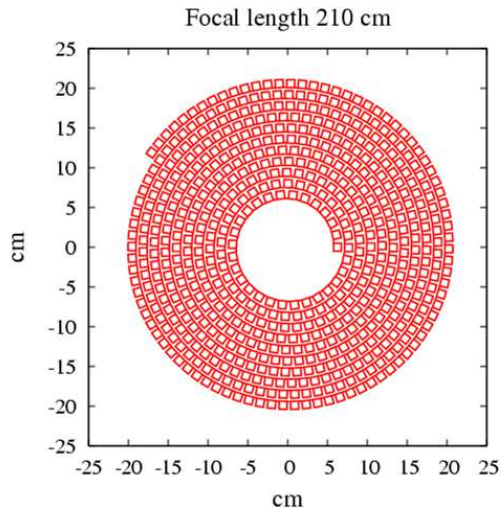, width=7cm}
  \end{center}
\caption[]{Disposition of the Cu (111) mosaic crystals in the lens according to
an Archimedes' spiral. This particular lens configuration refers to the prototype model now 
under development. \label{f:lens_pm}}
\end{figure}

\begin{figure}[ht]
  \begin{center}
    \epsfig{file=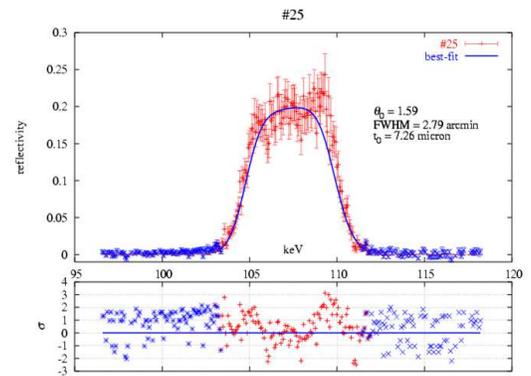, width=7cm}
  \end{center}
\caption[]{The measured reflectivity of a sample of  2 mm thick  Cu(111) produced and 
refined at ILL, Grenoble.  The used X--ray incident beam had a pencil-like geometry, with a cross
section of about $1 \times 1 $ mm$^2$, in order to test the uniformity of the mosaic
properties across the sample cross section.\label{f:tests}}
\end{figure}

Results of a feasibility study of a Laue lens with broad energy bandpass 
(typically 60--600 keV) have already been reported 
(\cite{Pisa04}). The technique adopted is Bragg diffraction from mosaic crystals. 
The geometrical configuration of the lens (see Fig.~\ref{f:lens}) is spherical, with sphere radius
equal to 2 times the focal length $FL$. The lens is covered by mosaic crystal
tiles with sizes as small as possible (e.g., $10 \times10$~mm$^2$ or less). 
The crystal tiles are positioned in the lens according to an Archimedes' spiral. 
In Fig.~\ref{f:lens_pm} we show an example of this crystal tile disposition for a lens prototype
which we are currently developing (see below). The Archimedes' spiral geometry
allows a smooth dependence of the lens effective area with energy. The size
of the lens and thus its effective area scales with the square of the focal length. 
The adopted material is Cu~(111), which we can produce with a mosaic structure
of angular spread $\ge 1$~arcmin (FWHM). Reflectivity test results (see example in Fig.~\ref{f:tests})
performed on samples of Cu~(111)  are consistent with the theoretical 
expectations (\cite{Pelli05}). A hard X--ray facility (LARIX) for testing the Laue lenses
is also being developed (\cite{Loffredo05}).

%
%

%
%

The development  of a Demonstration 
Model (DM) is now in progress. It consists of 30 tiles of Cu~(111) mosaic crystal with 5 arcmin (FWHM) 
mosaic spread and  $15 \times 15$~mm$^2$ cross section. The goal is to
establish the crystal assembling technique. 
The development of a prototype model (PM) with 500 crystals and 210 cm $FL$ is
already scheduled as the next step. The nominal energy band of the PM will be from 60 to 200 keV.
In addition to the laboratory tests, the PM is also expected to be tested 
in a balloon flight.

The focal plane detector required for the Laue lenses should have 
a high detection efficiency in the entire range of operation of the lenses 
(almost all focussed gamma--rays should be detected), 
a spatial resolution $\le 1$~mm, an energy resolution (FWHM) $<2$~keV at 500 keV for the study
of the nuclear and annihilation lines, and a 
detector sensitive to the photon linear polarization. The polarization sensitivity
is a key requirement given that most of the 
emission processes of gamma--ray radiation in the source classes mentioned in Section 1 are expected
to partially produce polarized photons.

%
%

\section{Two examples of multi-lens telescopes}
\label{s:examples}
\begin{figure}[ht]
  \begin{center}
    \epsfig{file=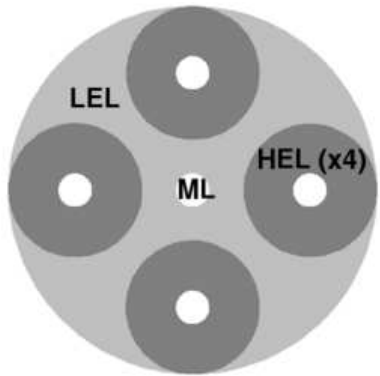, width=7cm}
  \end{center}
\caption[]{Top view of the multi-lens configuration 1, with 15 m focal length and a passband
from 60 to 600 keV. In the hole left free by the low energy lens, a multilayer telescope
is assumed, to extend the band down to $\sim 1$~keV.\label{f:conf1}}
\end{figure}

\begin{figure}[ht]
  \begin{center}
    \epsfig{file=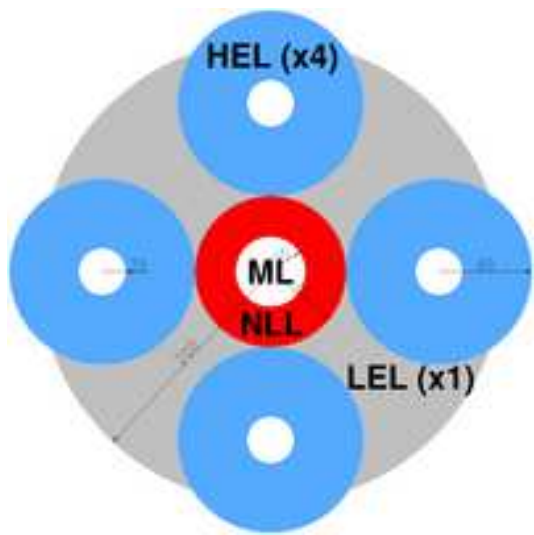, width=7cm}
  \end{center}
\caption[]{Top view of the multi-lens configuration 2, with a passband
from 60 to 900 keV. The LEL and the 4 HELs cover the band from 60 to 600 keV and 
have a focal length of 20 m; the lens devoted
to detect nuclear lines (NLL) in the range 450 to 540 keV and in 800--900 keV has a focal
length of 80 m. A multilayer 
telescope (ML), located in the hole left free by the NLL complements the lenses to extend 
the energy band down to $\sim 1$~keV.\label{f:conf2}}
\end{figure}

In order to show the expected capabilities of Laue lenses in the hard X-/soft gamma-ray
band, we report here two possible configurations of multi-lens telescopes.
\begin{enumerate}
\item 
{\bf Configuration 1} (see Fig.~\ref{f:conf1}) with a focal length of 15 m and a nominal  energy 
passband from 1 to 600 keV. It includes:
\begin{itemize}
\item 
1 Low Energy Lens (LEL) with a nominal passband from 54 to 200 keV;
\item
4 High Energy Lenses (HEL) with a nominal passband from 150 to 600 keV;
\item
In the hole left free from the LEL, a Wolter I Multi-Layer (ML) optics, confocal with the latter,
to cover the 1-80 keV energy band;
\end{itemize}
\item
{\bf Configuration 2} (see Fig.~\ref{f:conf2}) with a nominal  passband from 1 to 900 keV. It includes:
\begin{itemize}
\item
1 low energy lens (LEL) with a nominal 54-200 keV passband; 
\item
4 high energy lenses (HEL) with a nominal 150-600 keV passband; 
\item
1 Nuclear Line Lens (NLL) devoted to specifically study nuclear lines in the two energy bands: 450-540 and
 800-900 keV;
\item
In the hole left free from the NLL, a Wolter I Multi-Layer (ML) optics (not shown), confocal with the latter,
complements the lenses to cover the 1-80 keV energy band.
\end{itemize}
\end{enumerate}

In configuration 2, ML mirrors and LEL and HEL lenses are assumed to have a focal length of 20 m and
a mosaic spread of 2 arcmin, while the NLL lens is assumed to have a focal length of 80 m. The inconvenience
of the very high focal lengths ($>30$ m) is their requirement for a very small spread ($<1$~arcmin)
of the mosaic crystals, which is now much more difficult to produce. Multi--lens configurations
with one single focal length are being investigated. A ray tracing code devoted to study
the optical properties of the Laue lenses for offset incident beams is also in progress. 

%
%

%
%

\begin{figure}[ht]
  \begin{center}
    \epsfig{file=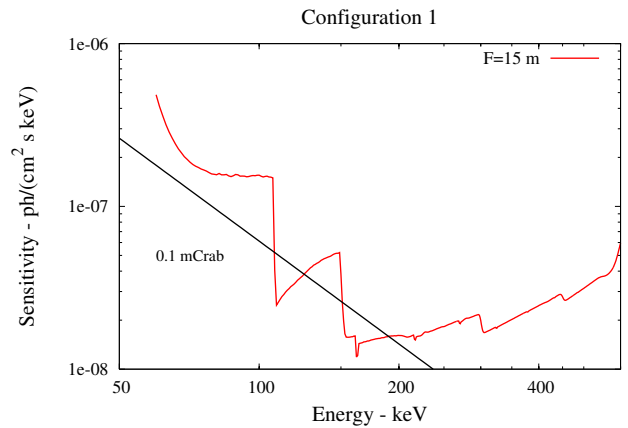,width=6cm,angle=-90}
  \end{center}
\caption[]{Expected $3\sigma$ sensitivity for a $10^6$ s exposure time in the case of Configuration 1.
The straight line shows the level of a 0.1 mCrab-like spectrum. A detection efficiency of $\approx 1$ 
is assumed. \label{f:sens1}}
\end{figure}

\begin{figure}[ht]
  \begin{center}
    \epsfig{file=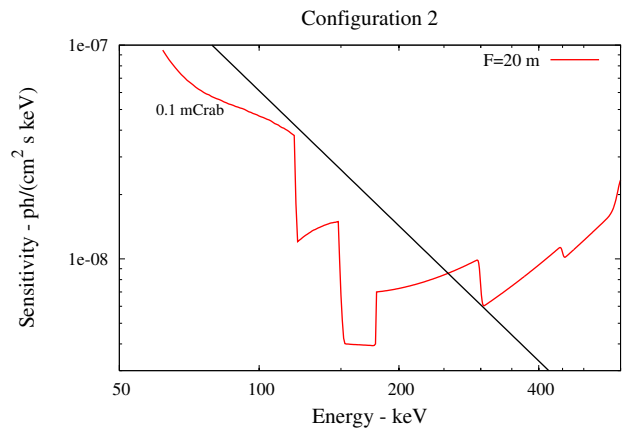,width=6cm,angle=-90}
  \end{center}
\caption[]{Expected $3\sigma$ sensitivity for a $10^6$ s exposure time in the case of Configuration 2.
The straight line shows the level of a 0.1 mCrab-like spectrum. A detection efficiency 
of $\approx 1$ is assumed. \label{f:sens2}}
\end{figure}

In Figs.~\ref{f:sens1} and \ref{f:sens2} we show the expected $3 \sigma$ sensitivity 
of  the  multi-lens Configurations 1 and 2 (only LEL and HEL), in which
a crystal thickness of 2~mm for the LEL and 4~mm for the HELs was assumed. The crystal thicknesses can be further
optimized to increase the lens sensitivity. Note that the LEL sensitivity can also be improved increasing 
its diameter.  The exposure time  is assumed to be $10^6$~s, while the band width is $\Delta E = E/2$. The 
detection efficiency is $\approx 1$.   

In configuration 2, the expected HEL sensitivity to a 511 keV annihilation line of 3 keV FWHM, is 
$F_{min} \approx 4\times 10^{-7}$~photons~cm$^{-2}$~s$^{-1}$. 
%
%

%
%

\section{Conclusions}

Laue lenses appear to be a promising approach in order to overcome significantly (even 
by 2 orders of magnitude, depending on energy) the sensitivity limitations of the current
generation of direct-viewing telescopes, with or without coded masks. We have shown
two examples of multi-lens configurations capable to cover the 60 to 600 keV band with
unprecedented sensitivity.

\begin{acknowledgements}
We acknowledge the financial support received from the Italian Space Agency ASI for this project in
2001 and 2002. We wish also to thank for their contribution to the DM development T. Franceschini
and S. Silvestri.
  
\end{acknowledgements}

\end{document}